\documentclass[times, 10pt,twocolumn]{article}
\usepackage{latex8}
\usepackage{times}
\usepackage[normalem]{ulem}
\usepackage{array}
\usepackage{multirow}
\usepackage{bbding}
\usepackage{array}
\usepackage{float}
\usepackage[linesnumbered,boxed]{algorithm2e}
\usepackage{amsmath}
\usepackage{times}
\usepackage{color}
\usepackage{authblk}
\usepackage{graphicx}
\usepackage{float}
\usepackage{url}

\begin{document}

\title{
BigDataBench: a Big Data Benchmark Suite from Internet Services }
\author[1,7]{\normalsize Lei Wang}
\author[1]{Jianfeng Zhan \thanks{The corresponding author is Jianfeng Zhan.}}
\author[1]{Chunjie Luo}
\author[1]{Yuqing Zhu}
\author[1]{Qiang Yang}
\author[2]{Yongqiang He}
\author[1]{Wanling Gao}
\author[1]{Zhen Jia}
\author[1]{Yingjie Shi}
\author[3]{Shujie Zhang}
\author[1]{Chen Zheng}
\author[1]{Gang Lu}
\author[4]{Kent Zhan}
\author[5]{Xiaona Li}
\author[6]{Bizhu Qiu}
\affil[1]{State Key Laboratory of Computer Architecture (Institute of Computing Technology, Chinese Academy of Sciences)
\authorcr \{wanglei\_2011, zhanjianfeng, luochunjie, zhuyuqing, yangqiang, gaowanling, jiazhen, shiyingjie, zhengchen, lugang\}@ict.ac.cn}
\affil[2]{Dropbox, yq@dropbox.com}
\affil[3]{Huawei,  shujie.zhang@huawei.com}
\affil[4]{Tencent, kentzhan@tencent.com}
\affil[5] {Baidu, lixiaona@baidu.com}
\affil[6]{Yahoo!,   qiubz@yahoo-inc.com}
\affil[7]{University of Chinese Academy of Sciences, China}
\date{}
\maketitle
\thispagestyle{empty}

\begin{abstract}
As architecture, systems, and data management communities pay greater attention to innovative big data  systems and architecture, the pressure of benchmarking and evaluating  these systems rises. However, the complexity, diversity, frequently changed workloads, and rapid evolution of big data systems raise great challenges in big data benchmarking.
Considering the broad use of big data systems, for the sake of fairness, big data benchmarks must include diversity of  data and workloads, which is the prerequisite for evaluating big data systems and architecture. 
Most of the state-of-the-art big data benchmarking efforts
target  evaluating specific types of applications or system
software stacks, and hence they are not qualified for serving
the purposes mentioned above.

This paper presents our joint research efforts on this issue with several industrial partners.
Our big data benchmark suite---\emph{BigDataBench}  not only covers broad application scenarios, but also includes diverse and representative data sets.  Currently,
we choose 19 big data benchmarks from dimensions of  application scenarios, operations/ algorithms, data types, data sources, software stacks, and application types, and they are comprehensive for  fairly measuring and evaluating big data systems and architecture. \emph{BigDataBench} is publicly available from the project home page {http://prof.ict.ac.cn/BigDataBench}.

Also, we comprehensively characterize 19 big data workloads included in \emph{BigDataBench} with varying data inputs. On a typical state-of-practice processor, Intel Xeon E5645,  we have the following observations: First, in comparison with the traditional benchmarks: including \emph{PARSEC}, \emph{HPCC},  and \emph{SPECCPU}, big data applications have very low \emph{operation intensity}, which measures the ratio of the total number of instructions divided by the total byte number of memory accesses; Second, the volume of data input has non-negligible impact on micro-architecture characteristics, which may impose challenges for simulation-based big data architecture research; Last but not least, corroborating the observations in \emph{CloudSuite} and \emph{DCBench} (which use smaller data inputs), we find that    the numbers of L1 instruction cache (L1I) \emph{misses per 1000 instructions} (in short, MPKI) of the big data applications are higher than in the traditional benchmarks;  also, we find that L3 caches are effective for the big data applications, corroborating the observation in \emph{DCBench}.
\end{abstract}
\section{Introduction}
Data explosion is an inevitable trend  as the world is connected more than ever. Data are  generated faster than ever, and to date about 2.5 quintillion bytes of data are created daily \cite{dataproduce}. This speed of data generation will continue in the coming years and is expected to increase at an exponential level, according to IDC's recent survey. The above fact gives birth to the widely circulated concept \emph{Big Data}.
But turning big data into insights or true treasure demands an in-depth extraction of their values, which heavily relies upon and hence boosts  deployments of  massive big data systems. 
As architecture, systems, and data management communities pay greater attention to innovative big data systems and architecture \cite{barroso2009datacenter, ferdman2011clearing}, \cite{zhan2012high}, the pressure of measuring, comparing, and evaluating  these systems rises \cite{bigbench}. Big data benchmarks are the foundation  of those efforts \cite{BigDataBenchISCA}.  However, the complexity, diversity, frequently changed workloads---so called workload churns \cite{barroso2009datacenter}, and rapid evolution of big data systems impose great challenges to big data benchmarking.

First, there are many classes of big data applications without comprehensive characterization. Even for internet service workloads, there are several important application domains, e.g., search engines, social networks, and e-commerce.  Meanwhile, the value of big data drives the emergence of innovative application domains.  The diversity of data and
workloads  needs comprehensive and continuous efforts on big data benchmarking. Second, most big data
applications are built on the basis of complex system software stacks, e.g., widely used Hadoop systems. However, there are not one-size-fits-all solutions \cite{wang2010transformer}, and hence big data system software stacks cover a broad spectrum.
Third, even if some big data applications are mature in terms of business and technology, customers, vendors, or researchers from academia or even different industry domains do not know enough about each other. The reason is that most internet service providers treat data, applications, and web access logs as business confidential, which prevents us from building benchmarks.

\begin{table*}[!t]
\centering
\caption{Comparison of Big Data Benchmarking Efforts} %
\label{BigDataBenchComp}%
\begin{footnotesize}
\begin{tabular}{|p{0.6in}|p{1.3in}|p{0.9in}|p{0.9in}|p{0.9in}|p{0.9in}|p{0.4in}|}
\hline
\textbf{Benchmark Efforts}&\textbf{Real-world data sets   (Data Set Number) } &\textbf{Data scalability (Volume, Veracity)} &\textbf{Workloads variety} &\textbf{Software stacks} & \textbf{Objects to Test} &\textbf{Status}\\\hline

HiBench \cite{hibench}& Unstructured text data (1)& Partial & Offline Analytics& Hadoop and Hive & Hadoop and Hive& Open \\
&&& Realtime Analytics & & &Source \\
\hline

BigBench \cite{bigbench}&None& N/A&Offline Analytics&DBMS and Hadoop&DBMS and Hadoop&Proposal\\
\hline

AMP &None & N/A&Realtime Analytics& Realtime analytic&Realtime analytic&Open \\
Benchmarks \cite{ampbig}&&&&systems&systems&Source\\
\hline

YCSB \cite{ycsb} &None &N/A &Online Services&NoSQL systems &NoSQL systems &Open Source \\
\hline

LinkBench \cite{armstrong2013linkbench} &Unstructured graph data (1) & Partial &Online Services& Graph database &Graph database &Open Source\\
\hline

CouldSuite & Unstructured text data (1)& Partial& Online Services&NoSQL systems,&Architectures &Open\\
\cite{ferdman2011clearing}&&&Offline Analytics&Hadoop, GraphLab&&Source\\
\hline

BigDataBench& Unstructured text data (1)& Total&  Online Services&NoSQL systems,& Systems and& Open  \\
& Semi-structured text data (1)&  & Offline Analytics&DBMS,&architecture; &Source\\
& Unstructured graph data (2)& &Realtime Analytics& Realtime Analytics &NoSQL systems; &\\
& Structured table data (1)& &&Offline Analytics &Different analytics &\\
& Semi-structured table data (1)&&&systems&systems&\\
\hline

\end{tabular}
\end{footnotesize}
\end{table*}

As summarized in Table \ref{BigDataBenchComp}, most of the state-of-the-art big data benchmark efforts target evaluating
specific types of applications or system software stacks, and hence fail to cover diversity of workloads and real-world data sets. However, considering the broad use of big data systems, \emph{for the sake of fairness, big data benchmarks must include diversity of workloads and data sets, which is the prerequisite for evaluating big data systems and architecture}. 
This paper presents our joint research efforts on big data benchmarking with several industrial partners.
Our methodology is from real systems,  covering not only broad application scenarios but also diverse and representative real-world data sets.
Since there are many emerging big data applications, we take an
incremental and iterative approach in stead of a top-down
approach. After investigating typical application domains of internet services---an important class of big data applications, we pay attention to investigating  workloads in three most important application domains according
to  widely acceptable metrics---the number of page views
and daily visitors, including search engine, e-commerce, and social network. To consider workload candidates, we make a tradeoff between choosing different types of applications: including online services, offline analytics, and realtime analytics.  In addition to workloads in three main application domains,  we include micro benchmarks for different data sources, "Cloud OLTP" workloads\footnote{OLTP is short for online transaction processing, referring to a class of information systems that facilitate and manage transaction-oriented applications with ACID (Atomicity, Consistency, Isolation, and Durability) support. Different from OLTP workloads, ¡±Cloud OLTP¡± workloads do not need ACID support.}, and relational queries workloads, since they are fundamental and widely used.  For three types of big data applications,  we  include both widely-used and state-of-the-art system software stacks.

From search engines, social networks, and e-commerce domains, six representative real-world data sets, whose varieties are reflected in two dimensions of data types and data sources, are collected, with the whole spectrum of data types including structured, semi-structured, and unstructured data. Currently, the included data sources are text, graph, and table data. Using these real data sets as the seed, the data generators \cite{BDGS} of \emph{BigDataBench}  generate synthetic data by scaling the seed data while keeping the data characteristics of raw data.
To date, we chose and developed nineteen big data benchmarks from dimensions of  application scenarios, operations/ algorithms, data types, data sources, software stacks, and application types.  We also plan to provide different implementations  using the other software stacks.  All the software code is available from \cite{opensource}. 

On a typical state-of-practice processor: Intel Xeon E5645, we comprehensively characterize nineteen big data workloads included in \emph{BigDataBench} with varying data inputs and have the following observation. First, in comparison with the traditional benchmarks: including \emph{HPCC}, \emph{PARSEC}, and \emph{SPECCPU}, the  floating point
operation intensity of\emph{ BigDataBench} is two orders of magnitude lower
than in  the traditional benchmarks. Though for the big data applications, the average ratio of integer instructions to floating point
instructions is about two orders of magnitude higher than in the traditional benchmarks, the average
integer operation intensity of the big data applications is still in the same order of magnitude like those of
the other benchmarks.
Second, we observe that the volume of data input has non-negligible impact on micro-architecture events. For the worst cases, the number of MIPS  (Million Instructions Per Second) of \emph{Grep} has a 2.9 times gap between the baseline and the 32X data volume; the number of L3 cache MPKI of \emph{K-means} has a 2.5 times gap between the baseline and the 32X data volume. This case may impose challenges for big data architecture research, since simulation-based approaches are widely used in architecture research and they are very time-consuming. Last but not least, corroborating the observations in \emph{CloudSuite} \cite{ferdman2011clearing} and \emph{DCBench} \cite{jiacharacterization} (which use smaller data inputs), we find that the numbers of L1I cache MPKI  of the big data applications are higher than in the traditional benchmarks.  We also find that L3 caches  are effective for the big data applications, corroborating the observation in \emph{DCBench} \cite{jiacharacterization}.

The rest of this paper is organized as follows. In Section 2, we discuss big data benchmarking requirements. Section 3 presents the related work. Section 4 summarizes our benchmarking methodology  and decisions----\emph{BigDataBench}.
Section 5 presents how to synthesize big data  while preserving characteristics of real-world data sets.  In Section 6,
we characterize \emph{BigDataBench}. Finally, we draw the conclusion in Section 7.
\Section{Big Data Benchmarking Requirements} \label{requirements}
This section discusses big data benchmarking requirements.

(1) Measuring and comparing big data systems and architecture.  First of all, the purpose of big data benchmarks is to measure, evaluate,  and compare big data systems and architecture in terms of user concerns, e.g.,  performance, energy efficiency, and cost effectiveness. Considering the broad use cases of big data systems, for the sake of fairness, a
big data benchmark suite candidate  must cover  not only broad application scenarios, but also diverse and representative real-world data sets.

(2) Being data-centric.  Big data are characterized  in
four dimensions called "4V" \cite{3v,IBM4vs}. \emph{Volume} means big data systems need to be able
to handle a large volume of data, e.g., PB.  \emph{Variety} refers to
the capability of processing data of different types, e.g., un-structured, semi-structured, structured data,  and different sources, e.g., text and graph data.
 \emph{Velocity} refers to the ability of dealing with regularly or irregularly refreshed data.
Additionally, a fourth V "\emph{veracity}" is added by IBM data scientists \cite{IBM4vs}. Veracity concerns the uncertainty of data, indicating that raw data characteristics  must be preserved in processing or synthesizing big data.

(3) Diverse and representative workloads. The rapid development of data volume and variety makes big data
applications increasingly diverse, and innovative application
domains are continuously emerging. Big data workloads chosen in the benchmark suite  should reflect diversity of application scenarios, and include workloads of different  types
so that the systems and architecture researchers could obtain the comprehensive
workload characteristics of big data, which provides useful guidance
for the systems design and optimization.

(4) Covering representative software stacks. Innovative software stacks are developed for specific user concerns. For examples, for online services, being latency-sensitivity is of vital importance. The influence of software stacks to big data workloads should not be neglected, so covering representative software stacks
is of great necessity for both systems and architecture research.

(5) State-of-the-art techniques.  In big data applications,
workloads change frequently. Meanwhile,  rapid evolution of big data
systems brings great opportunities for emerging techniques, and a
big data benchmark suite candidate should keep in pace with the improvements of
the underlying systems.  So a big data benchmark suite candidate  should include
 emerging techniques in different domains. In addition, it should be extensible for future  changes.

(6) Usability.  The complexity of big data systems in terms of application scenarios, data sets, workloads, and software stacks prevents ordinary users from easily using  big data benchmarks, so its usability is of great
importance. It is required that the benchmarks should be easy to deploy, configure,  and run, and the performance data should be easy to obtain.
\section{Related work}
We summarize the major benchmarking efforts for big data and compare them against BigDataBench in Table \ref{BigDataBenchComp}.
The focus of most of the state-of-the-art big data benchmark efforts is evaluating specific types of applications or system software stacks, and hence not qualified for measuring big data systems and architectures, which are widely used in broad application scenarios.

Pavlo et al.  \cite{calda} presented a micro benchmark for big data analytics. It compared Hadoop-based analytics to a row-based RDBMS system and a column-based RDBMS one.
It is the Spark \cite{spark} and Shark \cite{shark} systems that inspire the AMP Lab big data benchmarks \cite{ampbig}, which targets real-time analytic. This effort follows the benchmarking methodology in \cite{calda}. The benchmarks not only have a limited coverage of workloads, but also cover only table data. Its object under test is restricted to realtime analytics frameworks.
HiBench \cite{hibench} is a benchmark suite for Hadoop MapReduce and Hive.
It covers incomplete  data types and software stacks. GridMix \cite{gridmix} is a benchmark
specially designed for Hadoop MapReduce, which includes only micro benchmarks for text data.

Internet services players also try to develop their benchmark suites. Yahoo! released their cloud  benchmark specially for data storage systems, i.e, YCSB \cite{ycsb}. Having its root in cloud computing, YCSB is mainly for simple online service workloads----so called "Cloud OLTP" workloads. Armstrong et al. \cite{armstrong2013linkbench} characterized the social graph data and database workloads for Facebook's
social network, and presented the motivation, design, and implementation of LinkBench,
a database benchmark that reflects real-world database workloads for social network applications.
The TeraSort or GraySort benchmark \cite{sort} considers the performance and cost involved in sorting a large number of 100-byte records, and its workload is not sufficient to cover the various needs of big data processing. TPC-DS  is TPC's latest decision support benchmark, covering complex relational  queries for decision support. TPC-DS
handles some aspects of big data like volume and  velocity. Still, it lacks key data types
like semi-structured and unstructured data and key applications types like realtime analytics.  BigBench \cite{bigbench} is the recent effort towards designing big data benchmarks.  BigBench focuses on big data offline analytics, thus adopting TPC-DS as the basis and adding atop new data types like semi-/un-structured data, as well as non-relational workloads. Although BigBench has a complete coverage of data types, its object under test is DBMS and MapReduce systems that
claim to provide big data solutions, leading to  partial coverage of software stacks. Furthermore, currently, it is not open-source for easy usage and adoption.

Recently, architecture communities also proposed \emph{CloudSuite} \cite{ferdman2011clearing} for scale-out cloud workloads, and \emph{DCBench} \cite{jiacharacterization} for datacenter workloads. Those efforts include small data sets, e.g., only 4.5 GB for \emph{Naive Bayes} reported in \emph{CloudSuite} \cite{ferdman2011clearing}. Moreover, they fail to include diversity of real-world data sets and workloads. For example, for both \emph{CloudSuite} and \emph{DCBench}, realtime big data analytics workloads  are not included, while they are very important emerging big data workloads. Moreover, they paid little attention to how to generate diversity of  scalable big data sets (volume) while keeping their veracity.
\section{Our Benchmarking Methodology and Decisions}
This section presents our  methodology and decisions on \emph{BigDataBench}.
\subsection{Our Benchmarking Methodology}
\begin{figure*}[tb]
\centering
\includegraphics[width=6in,height=1.6in]{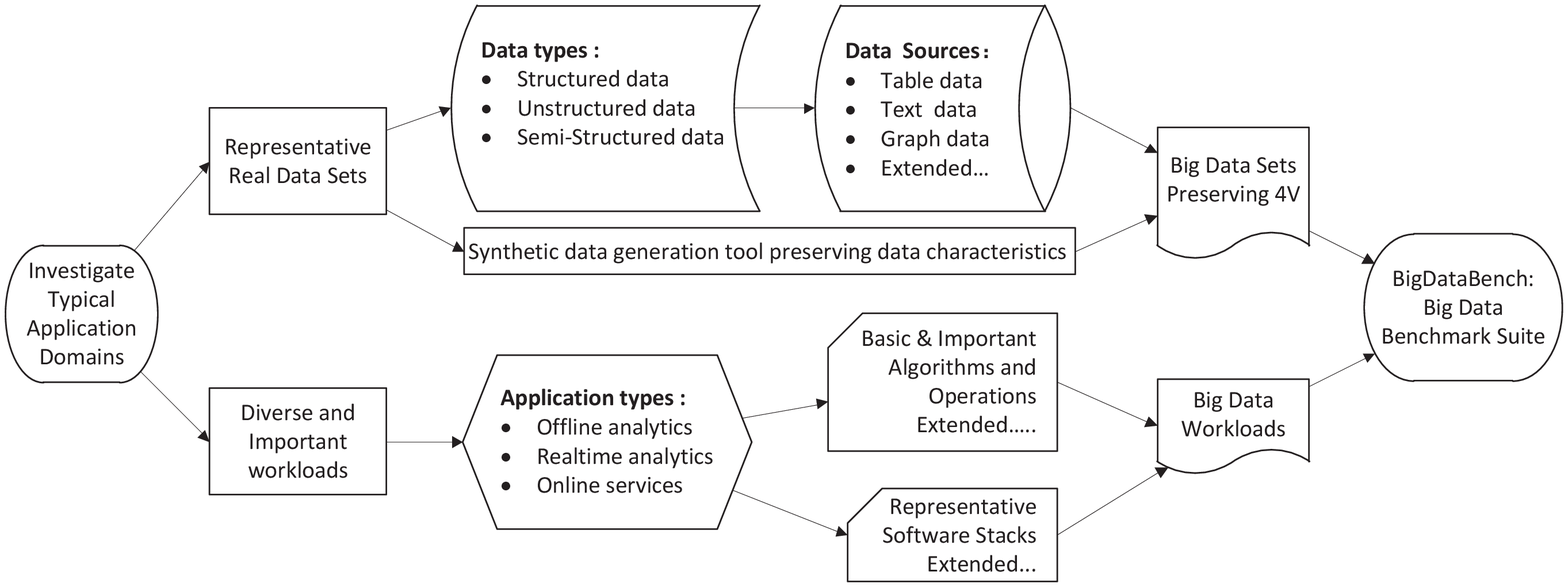}
\caption{BigDataBench Methodology.} 
\label{BigDataBenchM}
\end{figure*}

In this paper, we consider all the big data benchmarking requirements mentioned in Section \ref{requirements}  based on a solid-founded methodology as shown in
Figure. \ref{BigDataBenchM}.

As there are many emerging big data applications, we take an
incremental and iterative approach in stead of a top-down
approach. First of all, we investigate  the dominant application domains  of internet
services---an important class of big data applications according
to widely acceptable metrics---the number of page views
and daily visitors. According to the
analysis in \cite{Alexa}, the top three application domains are \emph{search
engines, social networks, and e-commerce},  taking up 80\%
page views of all the internet services in total.
And then,  we pay attention to typical  data
sets and big data workloads in the three application domains.

We consider data diversity in terms of both data types and data sources, and pay equal attention to  structured, semi-structured, and unstructured data. Further, we single out  three important data
sources in the dominant application domains of internet services, including
\emph{text data}, on which the maximum amount of analytics and  queries  are performed in search engines \cite{xi2011characterization},
\emph{graph data} (the maximum amount in social networks), and table
data (the maximum amount in e-commerce). Other important data sources, e.g., multimedia data,  will be
continuously added. Furthermore, we propose novel data generation
tools meeting with the requirements of data volume, variety, velocity, and veracity.

To cover diverse and representative workloads, we classify big data applications into three  types  from the users  perspective: online services, offline analytics, and realtime analytics. An online service is very latency-sensitive, and for each request, comparatively simple operations are performed for delivering  responses  to end users immediately. For offline analytics, complex computations are performed on big data with long latency. While for realtime analytics, end users want to obtain analytic results in an interactive manner. We pay equal attention to three application types. Furthermore, we choose typical workloads from two dimensions:
representative operations and algorithms from typical application scenarios,  widely-used and  state-of-the-art software stacks for three application types, respectively.

\subsection{Chosen  Data Sets}

As analyzed in the big data benchmarking requirements, the data sets
should be diverse and representative in terms of both data types and  sources.  After investigating three application domains, we collect six
representative \emph{real-world} data sets. Our chosen    data sets are diverse in three dimensions: data types, data sources,
and application domains. Table \ref{realdata} shows the
characteristics of six  real-world data sets.
The original data set sizes are not necessarily scaled to the hardware and software to be tested. We need to
scale the volume of the data sets while keeping their veracity, which we discuss  in Section \ref{data_generation}.

\begin{table}[htb]
\caption{The summary of real-world data sets.}\label{realdata}
\center
\begin{tabular}{|p{0.2in}|p{1.0in}|p{1.5in}|}

  \hline
  No.   &data sets            &   data size \\
  \hline
  1 & Wikipedia Entries    & 4,300,000 English articles\\
  \hline
  2 & Amazon Movie Reviews            & 7,911,684 reviews \\
  \hline
  3 &Google Web Graph              &   875713 nodes, 5105039 edges  \\
  \hline
  4 &Facebook Social Network          & 4039 nodes, 88234 edges \\
  \hline
  5 &E-commerce Transaction Data            & Table 1: 4 columns, 38658 rows. Table 2: 6 columns, 242735 rows  \\
  \hline
  6 & ProfSearch Person Resum\'{e}s       &  278956 resum\'{e}s\\
  \hline
\end{tabular}
\end{table}

\textbf{Wikipedia Entries} \cite{wikipedia}. The Wikipedia data set
is unstructured, consisting of  4,300,000 English articles.
Four workloads
use this data set, including \emph{Sort}, \emph{Grep},
\emph{WordCount} and \emph{Index}.

\textbf{Amazon Movie Reviews} \cite{amazonreview}. This data set is
semi-structured, consisting  of 7,911,684 reviews  on 889,176 movies
by 253,059 users. The data span from Aug 1997 to Oct 2012.  Two workloads use this data set, including \emph{Naive Bayes} for sentiment classification,  and \emph{Collaborative Filtering (in short, CF)}-- a typical recommendation algorithm.

\textbf{Google Web Graph} (Directed  graph)\cite{googleweb}. This data set is
unstructured, containing 875713  nodes representing web pages and
5105039 edges representing the links between web pages. This data
set is released  by Google as a part of Google Programming Contest.
We use it for \emph{PageRank}.

\textbf{Facebook Social Graph} (Undirected graph) \cite{facebookgraph}. This data set
contains 4039 nodes, which represent users, and  88234 edges, which
represent friendship between users. The data set is used for the graph mining workload-- \emph{Connected Components, in short (CC)}.

\textbf{E-commerce Transaction Data}. This data set is from an e-commerce
web site, which we keep anonymous by request. The data set is structured, consisting of two tables: ORDER
and order ITEM. The details are shown in Table \ref{schema_ABC}. This data set is used for the relational queries workloads.

\begin{table}
\caption{Schema of E-commerce Transaction Data }\label{schema_ABC}
\center
\begin{footnotesize}
\begin{tabular}{c c}
\textbf{ORDER} & \textbf{ITEM} \\
\hline
ORDER\_ID \textcolor{green}{INT} & ITEM\_ID \textcolor{green}{INT} \\ BUYER\_ID \textcolor{green}{INT} & ORDER\_ID \textcolor{green}{INT} \\CREATE\_DATE \textcolor{green}{DATE} & GOODS\_ID \textcolor{green}{INT} \\& GOODS\_NUMBER \textcolor{green}{NUMBER(10,2)} \\& GOODS\_PRICE \textcolor{green}{NUMBER(10,2)} \\& GOODS\_AMOUNT \textcolor{green}{NUMBER(14,6)} \\
\end{tabular}
\end{footnotesize}
\end{table}

\textbf{\emph{ProfSearch} Person Resum\'{e}s}. This data set is from a  vertical
search engine for scientists developed by ourselves,  and its  web site is \url{http://prof.ict.ac.cn}.  The data set is
semi-structured, consisting of 278956 resum\'{e}s automatically
extracted from 20,000,000 web pages of about 200 universities and research
institutions.
This data set is  used
for "Cloud OLTP" workloads.

We plan to add other real-world data sets to investigate the impact of different data sets on the same workloads.

\subsection{Chosen Workloads}
 We choose the
\emph{BigDataBench} workloads with the following considerations:  1) Paying equal attention to different types of applications: online service, real-time analytics, and offline analytics; 2)
Covering workloads in diverse and representative application scenarios ; 3) Including different data
sources: text, graph, and table data; 4) Covering the
representative big data software stacks.

In total, we choose 19 big data benchmarks.
Table \ref{benchmark_summary} presents \emph{BigDataBench} from perspectives of application scenarios, operations/ algorithms, data types, data sources, software stacks, and application types.
For some end users, they may just pay attention to big data application of a specific type.
For example, they want to perform an apples-to-apples comparison of software stacks for realtime analytics. They only need to choose benchmarks with the type of realtime analytics.
But if the users want to measure or compare big data systems and architecture,  we suggest they cover all benchmarks.
\doublerulesep 0.1pt
\begin{table*}[htb]
\center
\begin{footnotesize}
\caption{The Summary of BigDataBench.}\label{benchmark_summary}
\begin {tabular} {|c|c|c|p{0.8in}|p{0.5in}|c|}
\hline
Application& Application& Workloads&  Data&    Data& Software\\
Scenarios&          Type&    &       types&  source& Stacks\\
\hline

\multirow{4}{1in}{ Micro Benchmarks} & \multirow{4}{1in}{Offline Analytics} & Sort& \multirow{4}{*}{Unstructured}& \multirow{3}{*}{Text}&  \multirow{4}{1in}{Hadoop, Spark, MPI}\\
\cline{3-3}
& &Grep & & &\\
\cline{3-3}
& &WordCount & & & \\
\cline{3-3}
\cline{5-5}
& &BFS & &Graph& \\
\hline

\multirow{3}{1in}{Basic Datastore Operations ("Cloud OLTP" }& \multirow{3}{1in}{Online Service}& Read& \multirow{3}{*}{Semi-structured}& \multirow{3}{*}{Table}& \multirow{3}{1in} {Hbase, Cassandra, MongoDB, MySQL} \\
\cline{3-3}
& &Write&&&\\
\cline{3-3}
& &Scan&&&\\
\hline

\multirow{3}{1in}{Relational  Query} & \multirow{3}{1in}{Realtime Analytics} &Select Query
& \multirow{3}{*}{Structured}& \multirow{3}{*}{Table}&\multirow{3}{1in} {Impala, MySQL, Hive, Shark} \\
\cline{3-3}
& &Aggregate Query &&&\\
\cline{3-3}
& &Join Query &&&\\
\hline

\multirow{3}{1in}{Search Engine} & Online Services &Nutch Server& \multirow{3}{1in} {Un-structured}& \multirow{2}{1in}{Text}&\multirow{2}{1in}{Hadoop} \\
\cline{2-3}
& \multirow{2}{1in} {Offline Analytics}&                                      Index&    & &\\
\cline{3-3}
\cline{5-6}
& & PageRank& & Graph&Hadoop, Spark, MPI\\
\hline

\multirow{3}{1in}{Social Network} & Online Services &Olio Server& \multirow{3}{*}{Un-structured}& \multirow{3}{*}{Graph}
& Apache+MySQL \\
\cline{2-3}
\cline{6-6}
&\multirow{2}{1in} {Offline Analytics} &Kmeans &&&\multirow{2}{1in} {Hadoop, Spark, MPI}\\
\cline{3-3}
& &Connected Components (CC) && &\\
\hline

\multirow{3}{1in}{E-commerce} & Online Services &Rubis Server &Structured&
Table&Apache+JBoss+MySQL \\
\cline{2-6}
&\multirow{2}{1in} {Offline Analytics} &Collaborative Filtering (CF)& & &\multirow{2}{1in} {Hadoop, Spark, MPI}\\
\cline{3-3}
& &Naive Bayes &Semi-structured&Text&\\
\hline
\end{tabular}
\end{footnotesize}
\end{table*}

To cover diverse and representative workloads, we include important workloads from three important  application domains: search engines, social networks, and e-commence. In addition, we include micro benchmarks for different data sources, "Cloud OLTP" workloads, and relational queries workloads, since they are fundamental and pervasive.
The workload details are shown in the user manual available from \cite{opensource}.

For different types of big data applications, we also include widely-used and state-of-the-art system software stacks.  For example, for offline analytics, we include MapReduce, and MPI, which is widely used in HPC communities. We also include Spark, which is best for iterative computation. Spark  supports in-memory computing, letting it query data faster than disk-based engines like MapReduce-based systems. \emph{Most of the benchmarks in the current release \cite{opensource}, are implemented with Hadoop. But we plan to release other implementations, e.g., MPI, Spark}.
\section{Synthetic Data Generation Approaches and Tools} \label{data_generation}
How to obtain big data is an essential issue for big data benchmarking.
A natural idea to solve these problems is to generate synthetic data while keeping the significant features of real data. Margo Seltzer et al. \cite{seltzer1999case}  pointed that if we want to produce performance numbers that are meaningful in the context of real applications, we need use application-specific benchmarks.  Application-specific benchmarking  would need  application-specific data generation tools, which synthetically scale up real-world data sets while keeping their data characteristics \cite{tay2011data}. That is to say, for different data  types and sources, we need to propose different approaches to synthesizing big data.

Since the specific applications and data are diverse, the task of synthesizing big data on the basis of real-world data is nontrivial.
The data generation procedure in our benchmark suite is as follows:  First, we should have several  representative real-world data sets which are application-specific.
And then,  we  estimate the parameters of the data models using the real-world data.
Finally we generate synthetic data according to the data models and parameters, which are obtained from real-world data.

We develop \emph{Big Data Generator Suite} (in short, \emph{BDGS})--a comprehensive
tool--to generate synthetic big data preserving the 4V properties. The data generators  are designed for a wide class of application domains
(search engine, e-commence, and social network), and
will be extended for other application domains. We demonstrate its effectiveness
by developing data generators based on six real life data sets that cover three representative
data types (structured, semi-structured, and unstructured data), three
data sources (text, graph, and table). Each data generator can produce synthetic
data sets, and its data format conversion tools can transform these data sets into
an appropriate format capable of being used as the inputs of a specific workload.
Users can specify their preferred data size. In theory, the data size limit can only
be bounded by the storage size and the \emph{BDGS} parallelism in terms of the nodes
and its running time.  The details of generating  text, graph, and table data can be found at \cite{BDGS}.
\section{Workload Characterization Experiments}
In this section, we present our experiment configurations and methodology, the impact of the data volume on micro-architecture events, and workload characterization of big data benchmarks, respectively.
\subsection{Experiments Configurations and Methodology}
We run a series of workload characterization experiments using \emph{BigDataBench} to obtain  insights for architectural studies. Currently, we choose Hadoop as the basic software stack.
Above Hadoop, HBase and Nutch are also tested. Besides, MPICH2 and Rubis are deployed for understanding different workloads. In the near future, we will study the impact of different implementations on workload characterization using other analytic frameworks.

\begin{table}
\caption{Node configuration details of Xeon E5645}\label{hwconfigeration}
\center
\begin{tabular}{|c|c|c|c|}
  \hline
  \multicolumn{2}{|c|}{CPU Type} & \multicolumn{2}{|c|}{Intel CPU Core} \\ \hline
  \multicolumn{2}{|c|}{Intel \textregistered Xeon E5645}  &\multicolumn{2}{|c|}{6 cores@2.40G} \\ \hline
  \hline
L1 DCache &L1 ICache &L2 Cache &L3 Cache \\ \hline
6 $\times$ 32 KB& 6 $\times$ 32 KB&6 $\times$ 256 KB& 12MB \\ \hline
\end{tabular}
\end{table}

For the same big data application, the scale of the system running big data applications  is mainly decided by the size of data input. For the current experiments, the maximum data input is about 1 TB, and we deploy the big data workloads on the system with a matching scale---14 nodes. Please note that with our data generation tools in \emph{BigDataBench}, users can specify a larger data input size to scale up the real-world data, and hence need a larger system.
On our testbed, each node has two Xeon E5645 processors equipped with 16 GB memory and 8 TB disk. The detailed configuration of  each node is listed in Table \ref{hwconfigeration}. Please note that in the rest experiments, hyperthreading is enabled on our testbed. The operating system is Centos 5.5 with Linux kernel 2.6.34. The Hadoop distribution is 1.0.2 with Java version 1.6. The HBase, Hive,  MPICH2, Nutch, and Rubis distribution is  0.94.5, 0.9,  1.5, 1.1,  5.0, respectively. With regard to the input data, we vary the size from 32GB to 1TB for the analytics workloads. As it has large data complexity, the input data for the graph-related workloads like \emph{BFS}, \emph{CC}, and \emph{CF} workloads
are measured in terms of the set of vertices, while those of \emph{Index} and \emph{PageRank} workloads are in terms of Web pages. We also vary the request number from 100 requests per second to 3200 requests per second for all service workloads. Table \ref{WorkloadsEx} shows the workload summary.
\begin{table}[htb]
\center
\begin{footnotesize}
\caption{Workloads in experiments}
\label{WorkloadsEx}
\begin{tabular}{|l|l|l|l|}
  \hline
  ID         &   Workloads             & Software Stack   & Input size  \\
  \hline
  1       &   Sort             & Hadoop     & 32 $\times$(1,..,32) GB  data       \\
  \hline
  2       &   Grep             & Hadoop     & 32 $\times$(1,..,32)GB   data      \\
  \hline
  3       &   WordCount             & Hadoop     & 32 $\times$(1,..,32)GB data          \\
  \hline
  4       &   BFS             & MPI     & $2^{15} \times$(1,..,32)  vertex          \\
  \hline
  5       &   Read             & Hbase     & 32 $\times$(1,..,32) GB data        \\
  \hline
  6       &   Write             & Hbase     & 32 $\times$(1,..,32)GB data         \\
  \hline
  7       &   Scan             & Hbase     & 32 $\times$(1,..,32) GB data        \\
  \hline
  8       &   Select Query             & Hive     & 32 $\times$(1,..,32) GB data       \\
  \hline
  9       &   Aggregate Query             & Hive     & 32 $\times$(1,..,32)GB data          \\
  \hline
  10       &   Join Query             & Hive     & 32 $\times$(1,..,32)GB data         \\
  \hline
  11       &   Nutch server             & Hadoop     & 100  $\times$(1,..,32) req/s      \\
  \hline
  12       &   PageRank             & Hadoop     & $10^{6} \times$(1,..,32) pages        \\
  \hline
  13       &   Index             & Hadoop     & $10^{6} \times$(1,..,32) pages           \\
  \hline
  14       &   Olio Server             & MySQL     & 100 $\times$(1,..,32) req/s         \\
  \hline
  15       &   K-means             & Hadoop     & 32GB $\times$(1,..,32) data          \\
  \hline
  16       &  CC & Hadoop     & $2^{15}\times$(1,..,32) vertex          \\
  \hline
  17       &  Rubis Server             & MySQL     & 100 $\times$(1,..,32) req/s         \\
  \hline
  18       & CF & Hadoop     & $2^{15}\times$(1,..,32)  vertex          \\
  \hline
  19       & Naive Bayes & Hadoop     & 32 $\times$(1,..,32) GB data          \\
  \hline
  \end{tabular}
\end{footnotesize}
\end{table}

\subsubsection{Experiment Methodology}
Modern processors provide hardware performance counters to support micro-architecture level profiling.
We use Perf, a Linux profiling tool, to collect about 20 events whose numbers and unit masks can be found in the Intel Developer's Manual. In addition, we access the proc file system to collect OS-level performance data. We collect performance data after a ramp up period, which is about 30 seconds.

\subsubsection{Metrics} \label{Big_data_metrics}
Two categories of metrics are used for evaluation. The first category of metrics  are  \emph{user-perceivable metrics}, which can be conveniently observed and understood by users. The second ones are \emph{architectural metrics}, which are  mainly  exploited by architecture research. In the first category of metrics, we choose three measuring units for different workloads, respectively. The number of processed requests per second (\emph{RPS} in short) is used to measure the throughput of online service workloads. In addition, we also care about  latency. The number of operations per second (\emph{OPS} in short) is used to evaluate "Cloud OLTP" workloads. And, the data processed per second (\emph{DPS} in short) is used for analytic workloads\cite{luo2012cloudrank}. \emph{DPS} is defined as the input data size divided by the total processing time. 
In comparison with the metrics like the processed jobs or tasks per time unit, \emph{DPS} is much more relevant to the data processing capability of the system which users concern\cite{luo2012cloudrank}. The second category is chosen to compare performances under different workloads. Though the user-perceivable metrics can help evaluating different workloads in the same category, it is impossible to compare performances of workloads from different categories, e.g., a database server and a MapReduce workload. Hence, the uniform architecture metrics are necessary. Since no heterogeneous platform is involved in the experiments, we choose  the widely accepted performance metrics  in the architecture research, e.g., MIPS, and cache MPKI.

\subsubsection{The Other Benchmarks Setup}
For \emph{SPEC CPU2006}, we run the official applications with the first reference input, and report the average results into two groups:  integer benchmarks (\emph{SPECINT}) and floating point benchmarks (\emph{SPECFP}).  \emph{HPCC} is a representative HPC benchmark suite, and we run \emph{HPCC} with  version  1.4.  We run all  seven benchmarks, including \emph{HPL}, \emph{STREAM}, \emph{PTRANS}, \emph{RandomAccess}, \emph{DGEMM}, \emph{FFT}, and \emph{COMM}. \emph{PARSEC} is a benchmark suite composed of multi-threaded programs, and we deploy \emph{PARSEC} 3.0 Beta Release. We run all  12 benchmarks with native input data sets and use gcc with  version 4.1.2 to compile them.

\subsection{The Implication of Data Volume for Architecture Research}

Intuitively, data input should be highly relevant to big data workloads characterization. Specifically, the size of data input should be relevant to micro-architectural characteristics.
In this subsection, we pay attention to an important issue---\emph{what amount of data qualifies for being called big data from a perspective of workload characterization}? This issue is very important and interesting, because simulation is the basic approach  for architecture research, but it  is  very time-consuming. Bigger input data size would significantly increase the run time of a program, especially on the simulation platform. If there is no obvious difference between large and small  data inputs in terms of micro-architectural events, the simulation-based approaches using small data sets  can still be valid for architecture research. 
As different workloads have different input data types and sizes, we set the minimum data scales in Table \ref{WorkloadsEx} as the baseline data inputs, e.g., 32 GB for \emph{Sort}, 1000000 pages for \emph{PageRank}, and 100 requests per second for \emph{Nutch Server}. On the baseline data input, we scale up the data size by 4, 8, 16 and 32 times, respectively.

\begin{figure}[!t]
\centering
\includegraphics[scale=0.6]{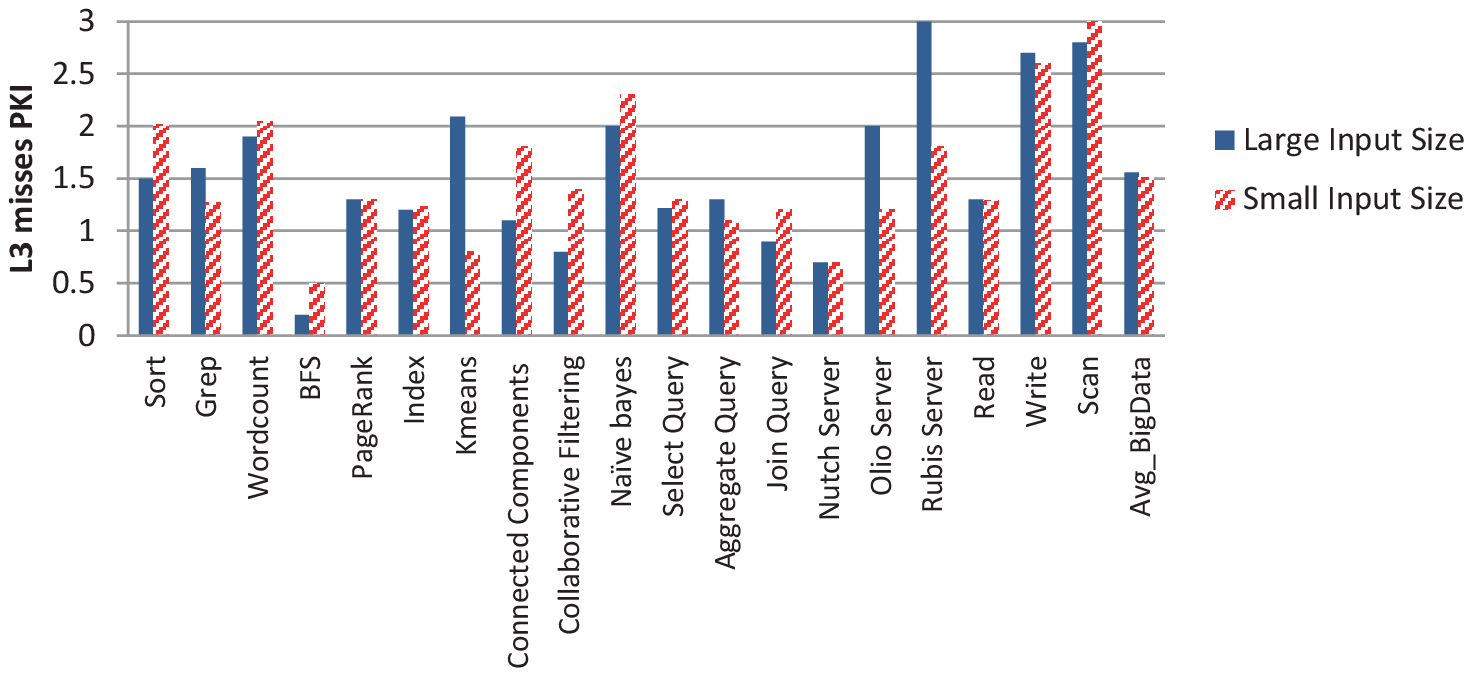}
\caption{L3 cache MPKI of different configurations in big data workloads.} 
\label{L3diff}
\end{figure}

\begin{figure*}[!t]
\centering
\includegraphics[scale=0.8]{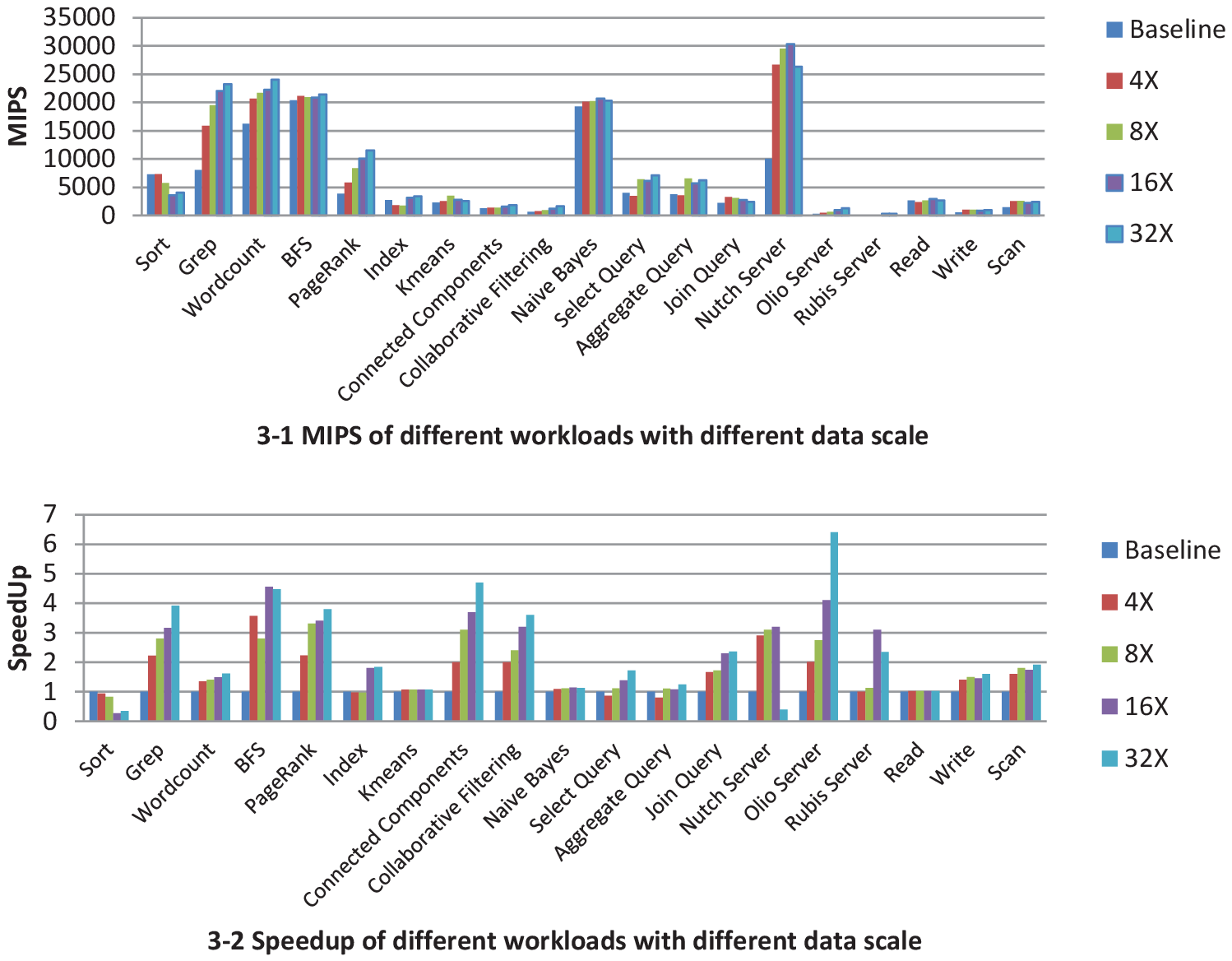}
\caption{Performance data vary with different data input sizes.} 
\label{Metrics}
\end{figure*}

There are hundreds of micro-architectural events in modern processors. For better readability and data presentation,   we only report the numbers of MIPS and L3 cache MPKI. 
Figure \ref{Metrics}-1 demonstrates MIPS numbers of each workload with different data scales. From Figure \ref{Metrics}-1, we find that for different workloads, the instruction executing behaviors exhibit different trends as the data volume increases. For example, MIPS numbers of  \emph{Grep} and \emph{WordCount} increase after the 16 times baseline, while for some other workloads, they tend to be stable  after the data volume increases to certain thresholds. The cache behavior metrics also exhibit a similar phenomenon as the MIPS metric does.  As important as L3 cache misses are--a single one can cause hundreds of cycles of latency--we track this value for different workloads under different configurations. In Figure \ref{L3diff}, for a workload,  we call  the data input on which the system achieves the best performance as  the \emph{large input} for a workload, and the baseline as the \emph{small input}. From Figure \ref{L3diff}, we can see that some workloads have lower number of L3 cache MPKI on the large configuration, e.g., \emph{Sort}, while some have higher number of L3 cache MPKI on the large configuration, e.g., \emph{Grep}. There are the other workloads showing no obvious difference under the two configurations, e.g., \emph{Index}. \emph{K-means}  has the largest difference under the two configurations, and the number of L3 cache MPKI is  0.8 and 2 for the small and large data inputs, respectively, which shows different data inputs can result in significantly different cache performance evaluation results.

To help understand the micro-architecture events, we also stress test  the cluster system with increasing data scale, and report the user-perceivable performance number as mentioned in Section \ref{Big_data_metrics}. Because workloads in Table \ref{WorkloadsEx} have different metrics,
we set the performance number  from the  experiments with the baseline data inputs as the baseline, and then normalize the collected results for each workload with varying data inputs  over the baseline number. For example, the performance number  of \emph{Workload A} for the baseline input is $x$ and that for 4 $\times$ baseline input is $y$, and then for \emph{Workload A} we normalize the performance number for  the baseline input and  4 $\times$ baseline input  as one  and  ($y \div x$), respectively.
Figure  \ref{Metrics}-2 reports the normalized performance numbers of each \emph{BigDataBench} workload with different data volumes.
Please note that the performance of \emph{Sort} degrades with increased data size in Figure  \ref{Metrics}-2 because \emph{Sort} is an I/O intensive workload when the memory cannot hold all its input data. Besides, the larger data sizes demand  more I/O operations like shuffling and disk accesses. Worse still, the network communication involved in data shuffling causes congestion, thus impairing performance.

[\textbf{Lessons Learned}].  As the above figures show, we find that different big data workloads have different performance trends as the data scale increases. This is the reason we believe that the workloads that only cover a specific application scenario  are not sufficient to evaluate big data systems and architecture. Second, architectural metrics are closely related to input data volumes and vary for different workloads, and  data volume has non-negligible impact on workload characterization.  For example, the MIPS number of \emph{Grep} has a 2.9 times  gap between the baseline and 32X data volume; the L3 cache MPKI of \emph{K-means} has a 2.5 times  gap between the baseline and 32X data volume.
This result implies that using  only simple applications with small data sets is not sufficient for big data systems and architecture research, which may impose great challenges.

\subsection{Workload Characterization}
This section mainly focuses on characterizing  operation intensity and cache behaviors of big data workloads.

\subsubsection{Measuring operation intensity.}

\begin{figure}[!b]
\centering
\includegraphics[scale=0.55]{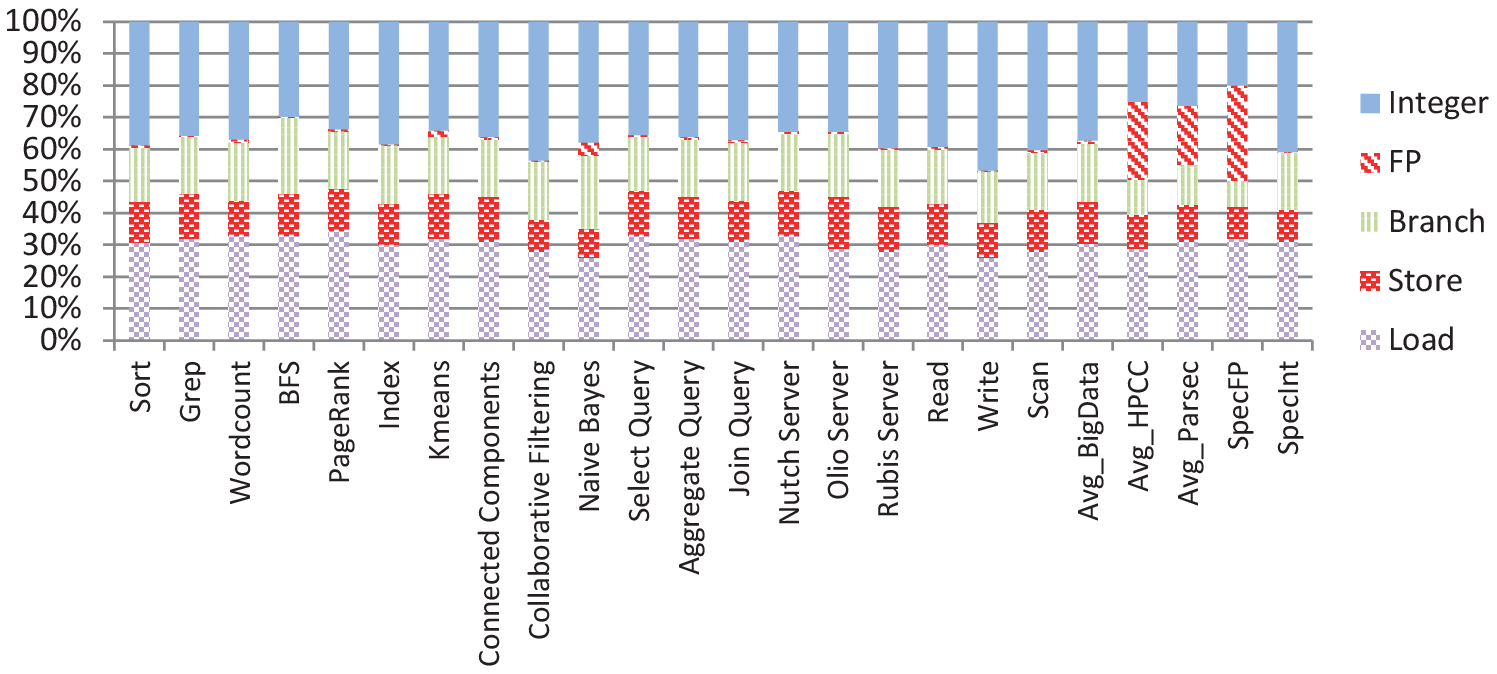}
\caption{Instruction Breakdown.} 
\label{InsMix}
\end{figure}

In order to characterize  instruction behaviors, first, we breakdown the execution instructions. As shown in Figure \ref{InsMix}, big data workloads have the distinct feature that the ratio of integer instructions to floating-point
instructions is very high. On Intel Xeon E5645, the average ratio is 75. The maximum is 179 (\emph{Grep}), and the minimum is 10 (\emph{Bayes}). For comparison, these ratios for \emph{PARSEC},\emph{ HPCC} and \emph{SPECFP} are very low, on the average 1.4, 1.0, and 0.67, respectively. The ratio for SPECINT (on the average 409) is  the highest,  because it is intended to  evaluate integer operations of processors. From this perspective, we can  conclude that the big data workloads  significantly differ from the traditional benchmarks like \emph{HPCC},  \emph{PARSEC}, and \emph{SPECCFP}. \emph{Please note that the reported numbers may deviate across different processors}. For example, Intel processors uses different generations of SSE (Streaming SIMD Extensions), which introduces both scalar and packed floating point instructions.

Furthermore, for each workload, we calculate \emph{the ratio of computation to memory access} to measure the operation intensity.  \emph{Floating point or integer operation intensity} is defined as the total number of (floating point or integer) instructions divided by the total number of memory accesses in terms of bytes in a run of the workload \cite{williams2009roofline}.
For example, in a run of \emph{program A}, it  has \emph{n} floating point instructions and \emph{m}   bytes of memory accesses, so the operation intensity of \emph{program A} is ($n \div m$). Since the memory hierarchy would impact the memory access performance significantly,
for comparison, we report experiments on two
state-of-practice processors: the Xeon E5310 and the Xeon E5460, respectively.
The Xeon E5310 is equipped with only two levels of caches, while the Xeon E5645 is equipped with
three levels of caches. The configuration of the Xeon E5310 is shown in Table. \ref{gdconfigeration}.
\begin{table}[h]
\center
\caption{Configuration details of Xeon E5310.}\label{gdconfigeration}

\begin{tabular}{|c|c|c|c|}
  \hline
  \multicolumn{2}{|c|}{CPU Type} & \multicolumn{2}{|c|}{Intel CPU Core} \\ \hline
  \multicolumn{2}{|c|}{Intel \textregistered Xeon E5310}  &\multicolumn{2}{|c|}{4 cores@1.60G} \\ \hline
  \hline
L1 DCache &L1 ICache &L2 Cache &L3 Cache \\ \hline
4 $\times$ 32 KB& 4 $\times$ 32 KB&2 $\times$ 4MB& None \\ \hline
\end{tabular}
\end{table}

\begin{figure}[!t]
\centering
\includegraphics[scale=0.60]{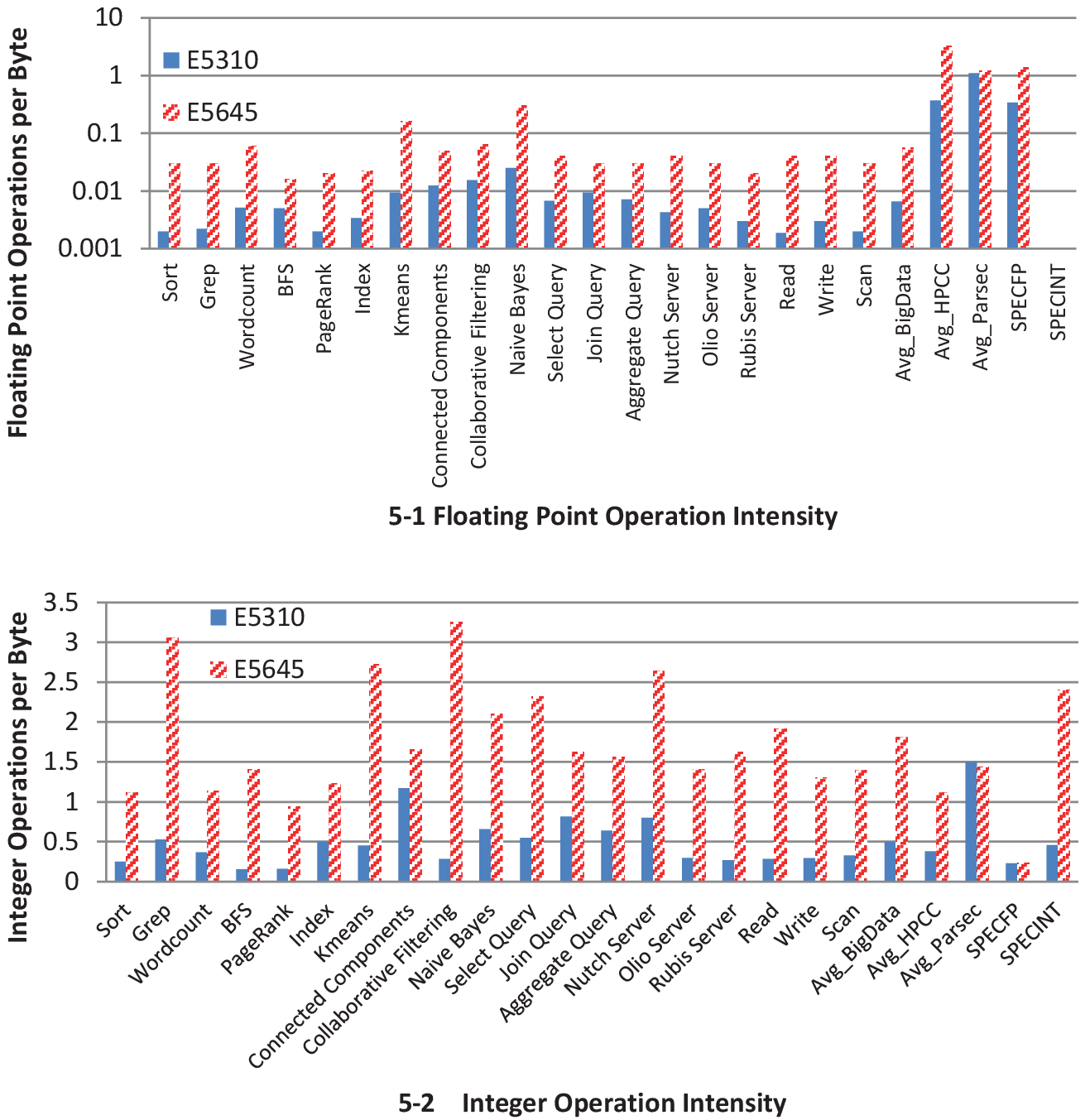}
\caption{Operation Intensity on Intel Xeon E5310 and E5645.} 
\label{Intensity}
\end{figure}

In Figure \ref{Intensity}-1, we can see that big data workloads have very low floating point operation intensities, and the average number of \emph{BigDataBench} is 0.007 on the Xeon E5310, and  0.05 on the Xeon E5645, respectively. However, the number of  \emph{PARSEC}, \emph{HPCC}, and \emph{SPCECFP} is higher as 1.1, 0.37, 0.34 on the Xeon E5310, and 1.2, 3.3, 1.4 on the Xeon E5645, respectively. \emph{SPECINT} is an exception with the number closing to 0. On the average, \emph{HPCC} and \emph{PARSEC} have high operation intensity because of their computing-intensive kernels, and \emph{SPECFP} has relatively high operation intensity for it is oriented for floating point operations. In summary, the floating point operation intensity of \emph{BigDataBench} is two orders of magnitude lower than in the traditional workloads on the Xeon E5310, and  Xeon E5645, respectively. The reason  the floating point operation intensity of \emph{BigDataBench} on E5645 is higher than on E5310 can be partly explained by the fact that  L3 caches are effective in decreasing the memory access traffic, which will be further  analyzed in  next subsection.

Though the average ratio of integer instructions to floating-point ones of big data workloads is about two orders of magnitude larger than in the other benchmarks, the average integer operation intensity of big data workloads is in the same order of magnitude like those of the other benchmarks. As shown in Figure \ref{Intensity}-2,  the average integer operation intensity of \emph{BigDataBench},  \emph{PARSEC}, \emph{HPCC}, \emph{SPECFP} and \emph{SPECINT} is 0.5, 1.5, 0.38,  0.23, 0.46 on the Xeon E5310 and 1.8, 1.4, 1.1, 0.2, 2.4 on the Xeon E5645, respectively.

[\textbf{Lessons Learned}].
In conclusion,  we can say that in comparison with the traditional benchmarks, the big data workloads in \emph{BigDataBench} have low ratios of computation to memory accesses. The above phenomenon can be explained from two aspects. First,  big data processing heavily relies upon memory accesses. Second,  big data workloads must process large volume of data, and hence most big data workloads adopt simple algorithms with low computing complexity. In \emph{BigDataBench}, they range from \emph{O(n)} to \emph{O(n*lgn)}. In comparison,  most \emph{HPCC} or \emph{PARSEC} workloads have higher computing complexity, ranging from \emph{O(n*lgn)} to \emph{O($n^3$)}.
We can make the conclusion that big data workloads have higher demand for  data movements than instruction executions. The state-of-practice processor is not  efficient for big data workloads. Rather, we believe that for these workloads the floating-point unit is over-provisioned.

\subsubsection{Memory Hierarchy Analysis.}
\begin{figure*}[!htb]
\centering
\includegraphics[scale=0.70]{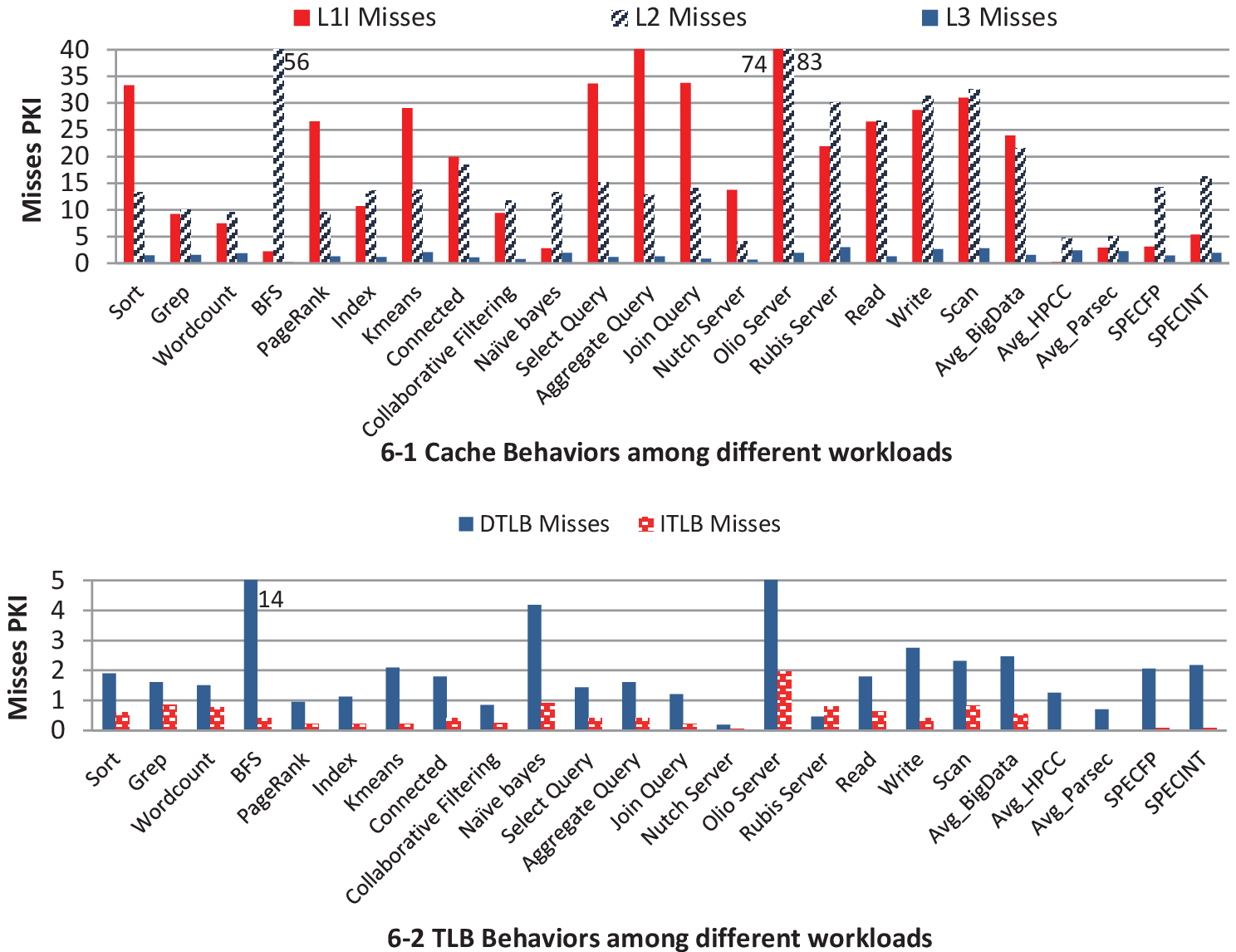}
\caption{Memory hierarchy behaviors among different workloads.} 
\label{BigDataCache}
\end{figure*}

 Finally, we show the operation intensity of the big data workloads is low, and we want to further investigate their  cache behaviors. In this subsection, we report L3 cache MPKI, L2 cache MPKI, L1 instruction MPKI, instruction TLB MPKI,  and data TLB MPKI, respectively. We leave out L1D cache MPKI because its miss penalty can be hidden by the
out-of-order  pipeline.   From Figure \ref{BigDataCache}, we observe that the cache behaviors of BigDataBench have three significant differences from the traditional benchmarks as follows:

First, the average L1I cache MPKI of \emph{BigDataBench}  is at least four times higher than in the traditional benchmarks. The average L1I cache MPKI of \emph{BigDataBench} is 23, while that of \emph{HPCC}, \emph{PARSEC}, \emph{SPECFP}, and \emph{SPECINT} is 0.3, 2.9, 3.1, and 5.4, respectively. This observation corroborates the ones in CloudSuite and DCBench. The possible main factors leading to the high L1I cache MPKI are the huge code size  and deep software stack of the big data workloads. Second, the average L2 cache MPKI of \emph{BigDataBench} is higher than in the traditional workloads. The average L2 cache MPKI of \emph{BigDataBench} is 21, while that for \emph{HPCC}, \emph{PARSEC}, \emph{SPECFP}, and \emph{SPECINT} is 4.8, 5.1, 14, and 16, respectively. Among \emph{BigDataBench}, most of the online service workloads have the higher L2 cache MPKI (on the average, 40) except \emph{Nutch server} (4.1), while most of the (offline and realtime) analytics workloads have the lower L2 cache MPKI (on the average, 13) except \emph{BFS} (56). Third, the average L3 cache MPKI of \emph{BigDataBench} is 1.5, while the average number of  \emph{HPCC}, \emph{PARSEC}, \emph{SPECFP}, and \emph{SPECINT} is 2.4, 2.3, 1.4 and 1.9, respectively. This observation shows that the LLC (L3) caches of the processors (Xeon E5645) on our testbed are efficient for the big data workloads, corroborating the observation in \emph{DCBench}. The efficiency of L3 caches also can explain why the floating point intensity of \emph{BigDataBench} on the Xeon E5645 is higher than on  the Xeon E5310 (only two levels of  caches).


The TLB behaviors are  shown in Figure. \ref{BigDataCache}-2. First, the average number of ITLB MPKI of \emph{BigDataBench} is higher than in the traditional workloads. The average number of ITLB MPKI of \emph{BigDataBench}  is 0.54, while that of \emph{HPCC}, \emph{PARSEC}, \emph{SPECFP},  and \emph{SPECINT} is 0.006, 0.005, 0.06, and 0.08, respectively. The more ITLB MPKI of \emph{BigDataBench} may be caused by the complex third party libraries  and deep software stacks of the big data workloads. Second, the average number of DTLB MPKI of \emph{BigDataBench} is also higher than in the traditional workloads. The average number of DTLB MPKI of \emph{BigDataBench} is 2.5, while that of \emph{HPCC}, \emph{PARSEC}, \emph{SPECFP}, and \emph{SPECINT} is 1.2, 0.7, 2, and 2.1, respectively. And we can also find that the numbers of DTLB MPKI of the big data workloads range from 0.2 (\emph{Nutch server}) to 14 (\emph{BFS}). The diversity of DTLB behaviors reflects the data access patterns are diverse in big data workloads, which proves that diverse workloads should be included in big data benchmarks.

[\textbf{Lessons Learned}]. On a typical state-of-practice processor: Intel Xeon E5645, we find that L3 caches of the processor are efficient for the big data workloads, which indicates multi-core CPU design should pay more attention to area and energy efficiency of caches for the big data applications. The high number of  L1I cache MPKI implies that better L1I cache performance is demanded for the big data
workloads. We conjecture that  the deep software stacks of the big data workloads are the root causes of high frond-end stalls. We are planning  further investigation into this phenomenon by changing the software stacks under test, e.g., replacing MapReduce with MPI.

\section{Conclusion}
In this paper, we presented our joint research efforts with several industrial partners on big data benchmarking. Our methodology is from real systems, covering not only broad application scenarios but also diverse and  representative real-world  data sets. We proposed  an innovative data generation methodology and tool  to generate scalable volumes of big data keeping the 4\emph{V} properties. Last, we chose and developed nineteen big data benchmarks from dimensions of  application scenarios, operations/ algorithms, data types, data sources, software stacks, and application types. Also, we reported the workloads characterization  results of big data as follows: first, in comparison with the traditional benchmarks, the big data workloads have very low operation intensity.
Second, the volume of data input has non-negligible impact on micro-architecture characteristics of big data workloads, so architecture research using only simple applications and small data sets is not sufficient for big data
scenarios. Last but not least, on a typical state-of-practice processor: Intel Xeon E5645, we find that for the big data workloads the LLC of the processor is effective  and better L1I cache performance is demanded as the big data workloads suffer high L1I cache MPKI.

\section{Acknowledgements}
We are very grateful to the anonymous reviewers and our shepherd, Dr. Michael D. Powell. This work is supported by Chinese 973 projects (Grants No. 2011CB302502 and 2014CB340402), NSFC projects (Grants No. 60933003 and 61202075) and the BNSF
project (Grant No.4133081).

\bibliographystyle{latex8}
\bibliography{references}

\end{document}